\documentclass[aip,
 amsmath,amssymb,
reprint,%
]{revtex4-1}

\usepackage{graphicx}
\usepackage{dcolumn}
\usepackage{bm}

\usepackage[utf8]{inputenc}
\usepackage[T1]{fontenc}
\usepackage{mathptmx}
\usepackage{etoolbox}

\usepackage{xcolor}

\usepackage{soul}

\makeatletter
\def\@email#1#2{%
 \endgroup
 \patchcmd{\titleblock@produce}
  {\frontmatter@RRAPformat}
  {\frontmatter@RRAPformat{\produce@RRAP{*#1\href{mailto:#2}{#2}}}\frontmatter@RRAPformat}
  {}{}
}%
\makeatother
\begin{document}

\preprint{AIP/123-QED}

\title[The effect of demagnetization on the susceptibility of single-domain particles and assemblies]{The effect of demagnetization on the susceptibility of single-domain particles and assemblies}
\author{Mathias Zambach}
\affiliation{DTU Physics, Technical University of Denmark, 2800 Kgs. Lyngby, Denmark}

\author{Miriam Varón}
\affiliation{DTU Physics, Technical University of Denmark, 2800 Kgs. Lyngby, Denmark}

\author{Mads R. Almind}
\affiliation{DTU Physics, Technical University of Denmark, 2800 Kgs. Lyngby, Denmark}

\author{Matti Knaapila}
\affiliation{DTU Physics, Technical University of Denmark, 2800 Kgs. Lyngby, Denmark}
\affiliation{Department of Physics, Norwegian University of Science and Technology, 7491 Trondheim, Norway}

\author{Ziwei Ouyang}
\affiliation{DTU Electro, Technical University of Denmark, 2800 Kgs. Lyngby, Denmark}

\author{Marco Beleggia}
\affiliation{Department of Physics, University of Modena and Reggio Emilia, 41125 Modena, Italy}
\affiliation{DTU Nanolab, Technical University of Denmark, 2800 Kgs. Lyngby, Denmark}

\author{Cathrine Frandsen}%
 \email{fraca@fysik.dtu.dk}
\affiliation{DTU Physics, Technical University of Denmark, 2800 Kgs. Lyngby, Denmark}

\date{\today}

\begin{abstract}
According to the classical laws of magnetism, the shape of magnetically soft objects limits the effective susceptibility. For example, spherical soft magnets cannot display an effective susceptibility larger than 3. Although this is true for macroscopic multi-domain magnetic materials, we explain why magnetic nanoparticles in a single-domain state do not suffer from this limitation. For single-domain particles, the differences between demagnetization factors along principal axes are relevant and can influence susceptibility but do not limit the susceptibility to an upper value as in the case for multi-domain particles. We experimentally validated this result on spherical nanoparticles with varying diameter (9 to 150 nm) and varying volume fraction (0.1 to 47 vol\%). In agreement with our predictions, we measure single-domain particle susceptibilities largely above 3, in fact up to more than 250. Moreover, contrary to an existing model for assemblies of particles, we find that the susceptibility of materials composed of non-interacting single-domain particles in a non-magnetic matrix scales linearly with the volume fraction of particles. This implies that high susceptibilities (>100) are achievable for nanoparticle-based composites and is relevant for the design of magnetically soft materials that are operational at MHz-GHz frequencies with negligible power losses.
\end{abstract}

\maketitle

Magnetic susceptibility - the change of magnetization upon a change in applied field - is one of the main material parameters to consider when designing magnetic components. Some materials have very high intrinsic susceptibilities; for example, iron has it above 1000. However, according to the shape of the magnet, the demagnetizing field diverging from the magnetic poles can limit the effective susceptibility to a purely geometric effect. It is a well-known text book result that the effective susceptibility, $\chi_{\textrm{p,eff}}$, of a (multi-domain) particle is given as \begin{equation}
    \chi_{\textrm{p,eff}}
    =\frac{1}{N+1/\chi_{\textrm{p}}},
    \label{eq:1}
\end{equation}
where $N$ is the demagnetization factor of the particle along the applied field direction and $\chi_p$ is the intrinsic susceptibility of the particle. The effective susceptibility for magnetically soft materials where $\chi_p\rightarrow\infty$ is then at most 1/$N$, or 3 for magnetic spheres where $N=1/3$. Consequently, large effective susceptibilities can only be achieved with particles elongated and aligned along the applied field direction.

Equation \eqref{eq:1} is derived implicitly for multi-domain magnets. One may wonder whether Eq. \eqref{eq:1} holds also for single-domain particles. As these are always magnetically saturated, their susceptibility may not be limited by demagnetization effects in the same way. Along these lines, Skomski et al. (2010) \cite{Skomski2010a} found unusual demagnetization effects in the magnetization processes of nano-scale spheres (due to coherent magnetization rotation), but did not explore it further.

For ensembles of magnetic particles, the demagnetization correction is anticipated to be more complex due to the combined contributions from the shape of the individual particles and that of the sample \cite{breit1923a,bleaney1941a,Skomski2007,bjoerk2013a,kin2016a,Sanchez2017,mccann2022a}.
A global demagnetization factor that depends on the particle volume fraction, $f$, and the demagnetization factors of both spherical particles $N=1/3$ and sample shape $N_s$, has been proposed \cite{breit1923a,bleaney1941a,Skomski2007,bjoerk2013a,Sanchez2017,mccann2022a} 
\begin{equation}
     {N_{\textrm{c}} = \frac{1}{3} + f\left( N_{\textrm{s}} - \frac{1}{3}\right).}
    \label{eq:BahlDemag}
\end{equation}
Although this model may apply to ensembles of multi-domain particles \cite{bjoerk2013a}, its validity for single-domain particles is uncertain\cite{bjoerk2013a,mccann2022a}, especially if Eq. \eqref{eq:1} does not hold for single-domain particles. With Eq. \eqref{eq:BahlDemag}, the effective ensemble susceptibility will be at most $\approx$ 11.5 (assuming $f=0.74$ for close-packed spheres and $N_\mathrm{s}=0$). Higher susceptibilities have been obtained on samples of nanoparticles. For instance, susceptibility measurements on systems of single-domain nanoparticles of Fe \cite{kura2012a, kin2016a}, FeNi \cite{yatsugi2019a, FeNi3_Article_Lu,liu2005a,WILLIAMS2019125}, FeCo \cite{kura2014a, yang2018a}, and ferrites \cite{yun2014a, yun2016a} that aim for application as high-susceptibility material in e.g. inductor cores indicate that spherical single-domain particles may not be limited by demagnetization effects in the same way as multi-domain soft spherical magnets, although this is not discussed in these references. Equation \eqref{eq:BahlDemag} has been used for nanoparticle assemblies\cite{normile2016a}, but there is a lack of verification of its general applicability to nanoparticle systems\cite{mccann2022a} as well as a lack of alternative models.

The design of high-susceptibility materials is of great interest. For example, new magnetic composite materials containing single-domain magnetic particles are being investigated for sensors \cite{Bedanta2007,Bertacco2017a,Bertacco2018a} and micro-inductor cores in power electronics \cite{sullivan2013a,FeNi3_Article_Lu,yun2014a,yun2016a,yatsugi2019a,liu2005a,kura2012a,kura2014a,yang2018a,kin2016a,garnero2019a,hasegawa2009a,rowe2015a,Bima_Preprint,zambach2023b,Zambach2025-MaterialPaper}. 
For these applications, it is crucial to understand the effective magnetic susceptibility of single-domain magnetic particles and the composite materials containing them. 

This letter takes a fundamental approach to clarify the underlying physics {regarding demagnetisation} to the questions: i) is the susceptibility of single-domain particles limited by shape in the same way as for multi-domain particles? and ii) which demagnetization factor describes an ensemble of non-interacting spherical single-domain particles?\\

First, we note that the demagnetization field in single-domain particles differs from the multi-domain case: the magnetization is saturated at all times in the 0 K limit. Hence, the magnitude of the demagnetization field in single-domain particles does not increase due to an induced magnetization with applied field as it happens for multi-domain particles. Instead, the demagnetization field rotates with the saturated magnetization according to the demagnetization factors. This modifies the way demagnetization factors enter the effective susceptibility for single-domain particles compared to the case of multi-domain particles.

For both single-domain and multi-domain particles, the particle susceptibility tensor components, $\chi_{\textrm{p}_{ij}}$, and the effective particle susceptibility, $\chi_{\textrm{p,eff}}$, can be written as
\begin{equation}
    \chi_{\textrm{p}_{ij}} = \frac{\partial M_i}{\partial H_{\textrm{i}_{j}}} \quad \text{and} \quad \chi_{\textrm{p,eff}} = \frac{\partial ( \mathbf{M} \cdot \hat{\mathbf{H}})}{\partial {H}}.
    \label{eq:ChiDerivation}
\end{equation}
Here, $M_i$ is the component of the magnetization in the $i$-direction, ${H}_{\textrm{i}_j}$ is the internal field strength of the particle in the $j$ direction, ${H}$ is the applied field amplitude, and $\hat{\mathbf{H}}$ is the applied field direction unit vector. For uniformly magnetized particles, the demagnetization field can be described by the demagnetization tensor components, $N_{jk}$, such that the internal field component becomes 
\begin{equation}
    H_{\text{i}_j} = H_j + H_{\text{d}_j} = H_j - N_{jk}M_k.
    \label{eq:DemagFieldTensor}
\end{equation}
In Eq.\eqref{eq:DemagFieldTensor} and thereafter we assume summation over repeated indexes.

In the multi-domain case, the linear relationship is $M_j = \chi_{\text{p}}H_{\text{i}_j}$. Thus, $\chi_{\text{p}_{ij}} = \left( \delta_{ij}/\chi_{\text{p}} + N_{ij} \right)^{-1}$, such that the effective susceptibility for multi-domain particles is
\begin{equation}
    \chi_{\text{p,eff}}= \frac{h_x^2}{N_{xx} + 1/\chi_{\text{p}}} + \frac{h_y^2}{N_{yy} + 1/\chi_{\text{p}}} + \frac{h_z^2}{N_{zz} + 1/\chi_{\text{p}}}
    \label{eq:MultiEffectiveSusc}
\end{equation}
with $h_{x,y,z}$ being the direction cosines of the applied field, i.e. $\hat{\mathbf{H}} = (h_x,h_y,h_z)$. This expression is usually known in the format of Eq. \eqref{eq:1}, where $N$ is the demagnetization factor in the applied field direction. 

In case of a single-domain (blocked) particle, the magnetic moment will, without applied field, be aligned to the easy axis. 
We set the z-axis along this direction so that without applied field $\hat{\mathbf{M}} = \hat{\mathbf{z}}$. The magnetostatic energy for a uniformly magnetized ellipsoid with saturation magnetization $M_s$ is 
\begin{align}
    \frac{E}{K_{\text{d}} V} &=  N_{xx} m_x^2 + N_{yy} m_y^2 + N_{zz} m_z^2  - \mathcal{H} \left( \hat{\mathbf{M}}\cdot\hat{\mathbf{H}} \right),
    \label{eq:ESWShape}
\end{align}
with $m_{x,y,z}$ being the directional cosines of the magnetization, $\hat{\mathbf{M}} = (m_x,m_y,m_z)$, $V$ being the particle volume, and $K_{\text{d}} = \mu_0 M_s^2 / 2$, and $\mathcal{H}=2H/M_s$. In the absence of an applied field, Eq. \eqref{eq:ESWShape} can be set equal to the contribution from the magnetization along the z-axis, i.e. $E/(K_{\text{d}} V)=N_{zz}$. Using $m_z\approx1$ for small fields, we can write
\begin{align}
    0 &=\left( N_{xx} - N_{zz} \right) m_x^2 + \left( N_{yy} - N_{zz} \right) m_y^2 \nonumber\\
    &- 2\mathcal{H} \left( m_x h_x + m_y h_y + h_z\right).
    \label{eq:ESWEqu}
\end{align}
From \eqref{eq:ESWEqu}, we find the $x$ and $y$ components of the magnetization by setting the energy gradients with respect to $m_x$ and $m_y$ to zero, obtaining
\begin{align}
    m_x &= \frac{\mathcal{H} h_x}{N_{xx} - N_{zz}} & &\text{and} & m_y &= \frac{\mathcal{H} h_y}{N_{y} - N_{zz}}.
\end{align}
Using equation \eqref{eq:ChiDerivation} the effective susceptibility along the easy (z) axis for a blocked single-domain particle, when only considering shape anisotropy, is then
\begin{align}
    \chi_{\text{p,eff}} &= 
    \frac{h_x^2}{N_{xx}-N_{zz}} + \frac{h_y^2}{N_{yy}-N_{zz}}.
    \label{eq:ChiEffSD}
\end{align}
Noticeably, this shows that the effective susceptibility of single-domain particles depends on the \textit{difference} in demagnetization factors along the principal axes of the given shape. Along the same lines as above we find that the susceptibility perpendicular to the easy axis of a prolate particle, where the shape anisotropy easy axis is set parallel to the magneto-crystalline uniaxial anisotropy easy axis, is
\begin{align}
    \chi_{\textrm{p,eff}} &= \frac{\chi_{\textrm{p}}'}{1+\chi_{\textrm{p}}'\left( N_{ii} - N_{jj} \right)},
    \label{eq:ChiBeff_new}
\end{align}
with the perpendicular Stoner-Wohlfarth susceptibility $\chi_{\textrm{p}}'=\frac{\mu_0 M_{\textrm{s}}^2}{2 K_{\textrm{u}}}$ and the difference between demagnetization factors along the principal axis of the spheroid $N_{ii}-N_{jj}$. Thus, the shape anisotropy does not limit the effective susceptibility to $1/N$ as it does for the multi-domain case. In detail, Eqs. \eqref{eq:ChiEffSD} and \eqref{eq:ChiBeff_new} establish that for ellipsoidal single-domain particles, the susceptibility will only be limited due to demagnetization effects by $1/(N_{ii} - N_{jj})$, which has no upper limit, and for spherical single-domain particles, where $N_{xx}=N_{yy}=N_{zz}=1/3$, $\chi_{\text{p,eff}}$ is not 3, but diverges towards the particle susceptibility from other anisotropies.
Equation \eqref{eq:ChiEffSD} and \eqref{eq:ChiBeff_new} also reveal that, in clear contrast to the multi-domain case, spherical single-domain particles achieve the highest susceptibility.

For sufficiently small single-domain particles, thermal energy can induce superparamagnetism. The particle magnetization, while it remains single-domain, reverses at time scales shorter than the observation time. The characteristic time scale, $\tau$, for relaxation between easy directions, depends on the anisotropy energy relative to the thermal energy. The time-averaged magnetization and susceptibility can be found as for a paramagnetic ion (see, e.g., \cite{elfimova2019a,zambach2023b}). 

To clarify the effect of demagnetization on superparamagnetic particle susceptibility we write the demagnetization energy of a uniformly magnetized single-domain spheroid particle as
\begin{equation}
    E_{\textrm{H}_\textrm{d}} = -\frac{\mu_0}{2} \int_V \mathbf{M_{\textrm{s}}}\cdot\mathbf{H}_{\textrm{d}} \textrm{d}V = K_{\textrm{sh}} V \sin^2\Theta,
    \label{eq:Edemag}
\end{equation}
with the shape anisotropy constant $K_{\textrm{sh}}= K_{\text{d}}\left( N_{ii} - N_{jj} \right)$. Here, $N_{\textrm{ii}}$ and $N_{\textrm{jj}}$ are the demagnetization factors along the principal spheroid axes. $K_{\textrm{sh}}$ is positive (/negative) for prolate (/oblate) spheroids, respectively, and $\Theta$ is the angle between the magnetic moment and the longer (/shorter) principal spheroid axis.
We find the superparamagnetic particle susceptibility to be
\begin{align}
    \chi_{\textrm{p,spm}}(\theta_{\textrm{H}}) = \frac{\epsilon_{\textrm{M}}}{2}\left[\sin^2\theta_{\textrm{H}} + {R'}/{R}\left(3\cos^2\theta_{\textrm{H}}-1\right) \right]
    \label{eq:ChiAngle}
\end{align}
with 
\begin{equation}
    R' = \int_0^1 \zeta^2 \exp\left( \epsilon_{\textrm{k}} \zeta^2 \right) \textrm{d}\zeta \quad\textrm{and}\quad R = \int_0^1 \exp\left( \epsilon_{\textrm{k}} \zeta^2 \right) \textrm{d}\zeta.
    \label{eq:RRdiff}
\end{equation}
for a non-interacting uniaxial-anisotropy particle with the applied field at an angle $\theta_{\textrm{H}}$ to the anisotropy axis \cite{zambach2023b}.
The energy ratios are 
\begin{equation}
    \epsilon_{\textrm{k}} = \frac{K V}{k_B T},
     \quad\textrm{and}\quad 
    \epsilon_{\textrm{M}} = \frac{\mu_0 V M_{\textrm{s}}^2}{k_B T},
    \label{eq:Epsilon}
\end{equation}
with $K$ being the effective anisotropy; here $K = K_{\textrm{sh}}$.
If the length difference between the axes is larger than 5-10\%, then shape anisotropy dominates over magneto-crystalline anisotropy for most soft magnetic particle materials.
For uniaxial anisotropy particles, the susceptibility ranges from $\epsilon_{\textrm{M}}$ to 0 in the case of large anisotropy ($\epsilon_{\textrm{k}}\gg1$) as $R'/R$ goes towards 1. For low anisotropy ($\epsilon_{\textrm{k}}\ll1$), the limit of $R'/R$ is $1/3$ and the susceptibility of the uniaxial anisotropy particle goes towards the random case value $\epsilon_{\textrm{M}}/3$ for all $\theta_{\textrm{H}}$.
Hence $\chi_{\textrm{p,spm}}$ equals $\epsilon_{\textrm{M}}/3$ for a spherical particle ($K_{\textrm{sh}}=0$). Therefore, neither in the superparamagnetic single-domain case is the susceptibility limited to 3 by demagnetization effects for a spherical particle. If the shape anisotropy increases (for $K_{\textrm{sh}}$ small), demagnetization results in slightly larger susceptibility along the long axis and slightly lower susceptibility perpendicular to the long axis. Thus, demagnetization in superparamagnetic particles acts in a way similar to that for the blocked case: the differences in demagnetization factors along primary axes in the particle are of importance, but in some cases, the anisotropy improves $\chi_{\textrm{p,spm}}$. For non-interacting, randomly oriented superparamagnetic particles, one can average over all possible directions for Eq. \eqref{eq:ChiAngle} and thus recover the well known per particle susceptibility
\begin{align}
    \langle \chi_{\textrm{p,spm}} \rangle = \frac{\epsilon_{\textrm{M}}}{3} = \frac{\mu_0 V M_{\textrm{s}}^2}{3 k_B T}.
    \label{eq:ChiRandom}
\end{align}
Eq. \eqref{eq:ChiRandom} reveals that shape anisotropy is washed out in random ensembles of superparamagnetic particles, since all individual contributions to the susceptibility are averaged by the randomness of the particle axis with respect to the applied field direction.

To test our predictions that effective particle susceptibilities above $1/N$ are possible for single-domain particles, we investigated a series of dilute samples ($f$ = 0.5-2 vol\%) of close-to-spherical fcc Co particles with sizes (mean diameters) ranging between 9 and 140 nm on a porous Al$_2$O$_3$ support, see Figs. S1-S2. {The 9 nm Co particles were prepared by a polyol method and coated with oleic acid before being deposited on the support under inert conditions.} The {samples with 19-140 nm Co} particles {were prepared by wet-impregnation on the support followed by calcination/annealing at 1000 \textdegree C at different gas conditions for 3-8 hours and resulting} sizes were determined by Transmission Electron Microscopy (TEM){ post the heat treatment, see details in supplementary material.}
The powders were then filled into cylindrical sample holders for vibrating sample magnetometry (VSM) measurements. Packing of the powder samples was not assumed to induce particle aggregation or alignment, as low volume fractions (0.5-2 vol\%) of particles were used on solid support. The Co particles were then {further} reduced at 900 \textdegree C by Argon with 2.4\% H$_2$ in the {VSM} sample holder\cite{almind2023a} and subsequently measured along the sample diameter by VSM. The 9 nm particles were not heat treated. Minor and major hysteresis loops were measured at 200 \textdegree C, except for the 9 nm Co particles, which were measured at room temperature (25 \textdegree C), see Fig. S3-S7 in the supplementary material. The saturation magnetization of the particles was found to be close to the bulk value of Co (1.45 MA/m) in all samples. The coercivity of the samples was determined from major loops ($\pm$ 796 kA/m applied field) 
while effective sample susceptibility was found from minor loops ($\pm$ 4 kA/m applied field) as the slope of the total sample magnetization (moment per sample volume) per applied field. {Minor loops are to a large degree linear with ellipsoid shape, as seen from Fig. S6-S7 in the supplementary material. The 9 nm particles showed slightly S-shaped minor loops.} The sample susceptibility was then demagnetization corrected for the cylindrical sample shape \cite{beleggia2005a} and the particle susceptibility was found by dividing by the volume fraction of the particles.
The particle susceptibility of the 9 nm particles was 1/$T$-corrected to 200 \textdegree C.

\begin{figure}[t!]
\centerline{\includegraphics[width=20pc]{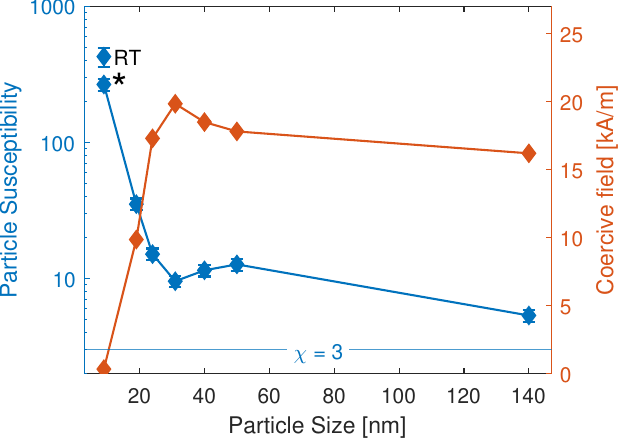}}
\caption{Initial particle susceptibility and coercive field of fcc Co particles of varying size on Al$_2$O$_3$ support at 200 \textdegree C after full reduction at 900 \textdegree C. The 9 nm particles were measured at room-temperature (RT) and $\chi$ was corrected as 1/$T$ to 200 \textdegree C (indicated by *).}
\label{fig:BlockedSusc}
\vspace*{-6pt}
\end{figure}

Figure \ref{fig:BlockedSusc} shows the initial particle susceptibility and coercivity for Co particles vs. size at 200 \textdegree C. The particle susceptibilities for the single domain particles (8-50 nm diameter) are above 9, proving that demagnetization does not limit particle susceptibility to 3. The susceptibility of the smallest particles is even greater than 250 at 200 \textdegree C (400 at room temperature). This sample does not have coercivity within the measurement uncertainty, which indicates superparamagnetism. The peak in coercive field indicates that particles with diameters between 30-60 nm are blocked. Co below 80 nm in diameter has been found to be single domain \cite{krishnan2010a}. 

In all, the measurements of Fig. \ref{fig:BlockedSusc} support that particle susceptibilities above 3 are possible for single-domain particles in both the blocked and superparamagnetic states.

Having derived the effect of demagnetization on single particle susceptibility, we turn to the topic of the susceptibility for materials containing single-domain magnetic particles in a non-magnetic matrix.
The most straightforward description would be a linear model \cite{elfimova2019a}, where the intrinsic susceptibility of the nanocomposite, $\chi_{\textrm{nc}}$, is simply the particle volume fraction, $f$, times the particle susceptibility
\begin{equation}
    \chi_{\textrm{nc}} = f \chi_{\textrm{p}}.
    \label{eq:LinearChi}
\end{equation}
This model is often used implicitly for ferromagnetic fluids or other dilute particle systems \cite{elfimova2019a,bai2018a,carrey2011a}. It seems readily adaptable to describe the susceptibility of a system of isolated single-domain particles where the particle susceptibility is not limited to ${1/N}$. 
However, the validity of the model has not yet been systematically investigated
for denser particle systems. Specifically, in cases where inter-particle interactions prevail, composite susceptibility may deviate from linear. 

Demagnetization correction for the nanocomposite sample shape should still be performed, even if Eq. \eqref{eq:LinearChi} is valid. That is, Eq. \eqref{eq:1} is applied to $\chi_{\textrm{nc}}$ of Eq. \eqref{eq:LinearChi} with the demagnetization factor $N=N_{\textrm{s}}$ in the applied field direction as 
\begin{equation}
    \chi_{\textrm{nc, eff}} = \frac{1}{N_s + 1/\chi_{\textrm{nc}}} = \frac{1}{N_s + 1/\left( f \chi_{\textrm{p}} \right)}.
    \label{eq:chi-ns-eff}
\end{equation}
If the model implied by Eqs. \eqref{eq:LinearChi} and \eqref{eq:chi-ns-eff} is valid for nanocomposites, then it precludes Eq. \eqref{eq:BahlDemag}, which corrects for both particle and sample shapes. Normile et al. \cite{normile2016a} have used the demagnetization factor from Eq. \eqref{eq:BahlDemag} to derive the intrinsic susceptibility of 9 nm maghemite particles in dense assemblies. However, they got large discrepancies in intrinsic susceptibilities (5 vs. 15) for similar particles in thin and thick disk samples with similar volume fractions of particles (50 vs 59 vol\%), suggesting that Eq. \eqref{eq:BahlDemag} may not be suitable for nanocomposites.

In order to test whether Eqs. \eqref{eq:LinearChi} and \eqref{eq:chi-ns-eff} are applicable to systems of single-domain particles we investigated a series of nanocomposites containing a varying volume fraction of spherical 11$\pm$3 nm $\gamma$-Fe$_2$O$_3$ (maghemite) particles fixed in a poly-vinyl alcohol polymer matrix\cite{Zambach2025-MaterialPaper}. The nanocomposite samples were cast as disks with a thickness of 100-400 $\mu$m and a diameter of 6 mm \cite{Zambach2025-MaterialPaper}. Small-angle neutron scattering confirmed near-spherical particle shape and particle size distributions obtained from TEM, as well as the absence of particle aggregation in the polymer matrix\cite{Zambach2025-MaterialPaper}.

The nanocomposites were measured by VSM with sample disk planes parallel to the applied field at 25 \textdegree C \cite{Zambach2025-MaterialPaper}. All nanocomposite samples showed the typical Langevin behavior for superparamagnetic nanoparticles that fits size distributions similar to the TEM results \cite{Zambach2025-MaterialPaper}. No detectable hysteresis was found with coercive fields below 4$\pm$8 A/m. Particle saturation magnetization of $M_{\textrm{s}} = 303$ kA/m was found from liquid samples for which the iron content had been quantified \cite{Zambach2025-MaterialPaper}. The magnetic particle volume content in the composites was calculated as the sample saturation magnetization divided by the particle saturation magnetization.

\begin{figure}[t!]
\centerline{\includegraphics[width=17.7pc]{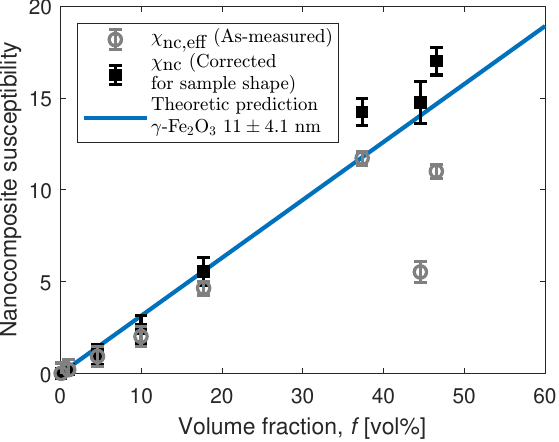}}
\caption{Susceptibility of nanocomposites containing randomly oriented {11}$\pm$3 nm diameter $\gamma$-Fe$_2$O$_3$ particles as function of particle volume together with theoretic prediction. $\chi_{\textrm{nc,eff}}$ (open circles) is the measured nanocomposite susceptibility and $\chi_{\textrm{nc}}$ (solid squares) is the measured susceptibility corrected for the sample shape as in Eq. \eqref{eq:chi-ns-eff}. Theoretic prediction based on Eqs. \eqref{eq:ChiRandom}-\eqref{eq:LinearChi}, with log-normal size distribution, $M_\textrm{s}=303$ kA/m, and $T=25$ \textdegree C.}
\label{fig:CompositeSusc}
\vspace*{-6pt}
\end{figure}

Figure \ref{fig:CompositeSusc} shows the susceptibilities of the nanocomposites obtained from VSM measurements. Here, $\chi_{\textrm{nc,eff}}$ (open circles) is given as-measured, and $\chi_{\textrm{nc}}$ (solid squares) is demagnetization corrected for nanocomposite sample shape (disk /flat cylinder), cf. Eq. \eqref{eq:chi-ns-eff}. The sample shape demagnetization factors are calculated according to Ref. \cite{beleggia2005a}. 
When demagnetization corrected for the sample shape, the nanocomposite material follows a close-to-linear trend for $\chi_{\textrm{nc}}$ in correspondence with Eq. \ref{eq:LinearChi}. 

For the nanocomposites in Fig. \ref{fig:CompositeSusc}, the corresponding intrinsic particle susceptibility $\chi_{\textrm{p}}$ (cf. Eq. \eqref{eq:LinearChi}) is consistently around 25 independent of the volume fraction (going from 22 up to 38 for the most concentrated samples
\cite{Zambach2025-MaterialPaper}). First, this confirms that  $\chi_{\textrm{p}}$ is not limited to 3 for spherical nanoparticles. Second, it exemplifies that Eq. \eqref{eq:BahlDemag} cannot hold, since such consistency in $\chi_{\textrm{p}}$ is not achievable by Eq. \eqref{eq:BahlDemag}.
Moreover, the high susceptibilities expose experimentally that the demagnetization factor in Eq. \eqref{eq:BahlDemag}, which corrects for both particle shape (as $1/3$) and sample shape, is not applicable. The measured nanocomposite susceptibilities of 11.0 - 11.7 (Fig. \ref{fig:CompositeSusc}), are too high to be explained by the demagnetization factor in Eq. \eqref{eq:BahlDemag}: $\chi_{\textrm{nc,eff}}$ could maximally be 5.6 for 46 vol\% of spherical nanoparticles, if using the demagnetization factor in Eq. \eqref{eq:BahlDemag}. Additionally, \textit{if} the nanocomposite susceptibilities of 4.7-11.7  measured here for the 17-46 vol\% samples (Fig. \ref{fig:CompositeSusc}) were demagnetization corrected for both the particle shape and the sample shape, as in Eqs. \eqref{eq:1}-\eqref{eq:BahlDemag}, then meaningless negative intrinsic susceptibilities, $\chi_{\textrm{nc}}$, of -5.1 to -9.4, are obtained, equivalent to overskewed hysteresis loops \cite{Skomski2010a}.

Overall, this shows that the demagnetization factor given in Eq. \eqref{eq:BahlDemag}, which may apply to composites of multi-domain particles \cite{bjoerk2013a}, is not applicable to systems of isolated single-domain nanoparticles. $\chi_{\textrm{nc,eff}}$ should be corrected for the sample shape, but not the spherical particle shape as $1/3$. 
Hence, to derive $\chi_{\textrm{nc}}$ from experimental measurements on nanocomposites containing isolated spherical single-domain particles, we suggest the following procedure: for a specific sample (e.g. a thin disk of nanocomposite), the effective susceptibility is measured in a certain direction and the measurement is demagnetization corrected for the sample shape only.

In general, influences of inter-particle interactions are a potential concern for the design of dense nanoparticle composites. In the literature, interactions have been reported to both increase and decrease the susceptibility of particles \cite{kin2016a,elfimova2019a,chantrell2001a,usov2020a}. In our test samples containing up to 46 vol\% $11\pm4.1$ nm $\gamma$-Fe$_2$O$_3$ particles, we observe some possible signs of inter-particle interactions with increasing volume fraction; cf. Eq. \eqref{eq:LinearChi} applies although the linearity of data is not perfect.
We estimate that the dipolar interaction between particle pairs in our system (1-46 vol\% {11}$\pm$3 nm  $\gamma$-Fe$_2$O$_3$) is on the scale of 
7-70$\times10^{-22}$ J, using  $\mu_0 m^2/(4\pi r_{\textrm{cc}}^3)$, where $r_{\textrm{cc}}$ is the mean center-center distance of the particles. This estimate does not take into account the fluctuating nature of particle magnetization and the fact that one particle has several neighbors, which may lead to increased/decreased interaction fields depending on the packing configurations of the particles \cite{var2013a, Durhuus2025a}. Compared to the anisotropy barrier ($KV$) which is in the range of 
170$\times10^{-22}$ J when using $K\approx25$ kJ/m$^3$, the interaction energy scale found for our system may barely be significant. This is also in agreement with the simulations in \cite{Durhuus2025a} for similar $\epsilon_{\textrm{k}}$ and $\epsilon_{\textrm{M}}$.
{For the low-density Co samples ($<2$ vol\%), the dipolar interactions between two particles are estimated to be on the order $1.6$ - $5800\times10^{-21}$ J using particle diameters of 9-140 nm. This interaction energy is one to two orders of magnitude lower than the anisotropy energy of the single particles.}\\

In conclusion, we explained why the effective single-domain particle susceptibility is not limited by demagnetization, neither for superparamagnetic nor for blocked single-domain particles. The difference between demagnetization factors along the applied field direction and the particle easy axis is of importance for single-domain particles but enters mainly like an extra (shape) anisotropy, and for spherical particles the susceptibility is not affected by particle demagnetization. It is noticeably that the near-spherical shape is the most favorable for obtaining a high effective susceptibility in single-domain particles, in contrast to multi-domain cases.
We have experimentally validated the improvement over the classical susceptibility limit by two nanocomposite systems. Spherical fcc Co particles of different sizes verify that neither superparamagnetic nor blocked single-domain particles have a particle susceptibility limited by demagnetization (i.e. $\chi_p \gg 3$). Spherical superparamagnetic $\gamma$-Fe$_2$O$_3$ particles in different volume fractions further confirm that the conventional particle demagnetization limit does not apply (i.e. $\chi_p \gg 3$) and conclude that the susceptibility of nanocomposites depends linearly on the particle volume fraction for non-interacting spherical nanoparticles (cf. Eq. \eqref{eq:LinearChi}). Consequently, Eq. \eqref{eq:BahlDemag} does not apply to spherical single-domain particles. The presented results are important for the design of new soft magnetic materials based on nanoparticles.

\section*{Supplementary Material}
The supplementary material contains TEM images, size analysis, heat treatment routines during synthesis, major and minor hysteresis loops, and comparison of particle susceptibility from major and minor hysteresis for the Co particles.

\begin{acknowledgments}
The authors thank Ron B. Goldfarb for providing helpful feedback to a previous version of this manuscript and the Independent Research Fund Denmark for financial support (project HiFMag, grant number 9041-00231A).
\end{acknowledgments}

\section*{Data Availability Statement}
The data that support the findings of this study are available from the corresponding author upon reasonable request.

\section*{References}
\bibliography{aipsamp}

\end{document}